\pdfoutput=1
\documentclass{article}

\usepackage[preprint]{neurips_2026}

\usepackage[utf8]{inputenc} 
\usepackage[T1]{fontenc}    
\usepackage{hyperref}       

\usepackage{url}            
\usepackage{booktabs}       
\usepackage{graphicx}       
\usepackage{placeins}       
\usepackage{amsmath}        
\usepackage{amsfonts}       
\usepackage{nicefrac}       
\usepackage{microtype}      
\usepackage[table, dvipsnames]{xcolor}         

\usepackage{tabularx}
\usepackage{booktabs}
\usepackage{array}
\usepackage{adjustbox}

\definecolor{myblue}{HTML}{0064E0}
\definecolor{myred}{HTML}{DC6F7B}
\usepackage{wrapfig} 
\hypersetup{
    colorlinks,
    linkcolor=myblue,
    citecolor=myblue,
    filecolor=Mulberry,
    urlcolor=RoyalBlue,
}

\title{Benign in Isolation, Harmful in Composition: Security Risks in Agent Skill Ecosystems}

\author{
    \textbf{Yi Xie}\textsuperscript{*,1}, 
    \textbf{Jiawei Du}\textsuperscript{*,2}, 
    \textbf{Yu Cheng}\textsuperscript{1,3}, 
    \textbf{Jiuan Zhou}\textsuperscript{1}, 
    \textbf{Zhaoxia Yin}\textsuperscript{\(\dagger\),1} \\ \textsuperscript{1}East China Normal University, Shanghai, China \\ 
    \textsuperscript{2}Centre for Frontier AI Research A*STAR, Singapore \\ 
    \textsuperscript{3}Shanghai Innovation Institute, Shanghai, China \\ 
    \small\textsuperscript{*}Equal contribution.
    \small\textsuperscript{\(\dagger\)}Corresponding author.
}

\begin{document}

\maketitle

\begin{abstract}
Skills are becoming the capability layer through which LLM agents turn plans into actions, but their use also introduces security risks, including data leakage, unauthorized operations, and tool misuse. Existing skill vetting typically reviews each skill in isolation, whereas real agent tasks often involve multiple skill invocations across a shared execution context. A risk therefore emerges: one skill’s output, trust signal, authorization cue, or side effect can be carried into a later skill invocation. We define this resulting gap as \textbf{SCR} (\textbf{S}kill \textbf{C}omposition \textbf{R}isk): skills that appear benign under isolated evaluation can become harmful along an activated composition path. \textbf{ SCR-Bench} (\textbf{S}kill \textbf{C}omposition \textbf{R}isk Bench) is introduced to evaluate this risk in controlled, sandboxed skill environments. Instead of relying on the textual intent or surface-level behavior of individual skills, SCR-Bench records downstream state changes and path-level outcomes across activated skill paths. SCR-Bench contains three sub-benchmarks: \textbf{SCR-CapFlow} for capability-flow composition, \textbf{SCR-TrustLift} for trust-transfer composition, and \textbf{SCR-AuthBlur} for authorization-confusion composition. Across SCR-Bench, composed skill paths expose risks that are largely absent under isolated evaluation: in SCR-CapFlow, ASR reaches 33.6\% under composition, compared with near-zero isolated baselines; in SCR-TrustLift, ASR reaches near saturation above 96.5\% on four of five backends; and in SCR-AuthBlur, the risky-approval rate increases by 71.8\% relative to the L0 isolated baseline under the L1 context setting. The experimental results reveal the core pattern of “Benign in Isolation, Harmful in Composition,” showing that agent skill security should be assessed at the level of activated paths rather than isolated artifacts. By introducing SCR and SCR-Bench, this work aims to support path-aware risk evaluation and defense in LLM agent skill ecosystems. Our benchmark is available at \url{https://github.com/saint-viperx/SCR_Bench}.
\end{abstract}

\begin{figure}[t]
  \centering
  \includegraphics[width=\linewidth]{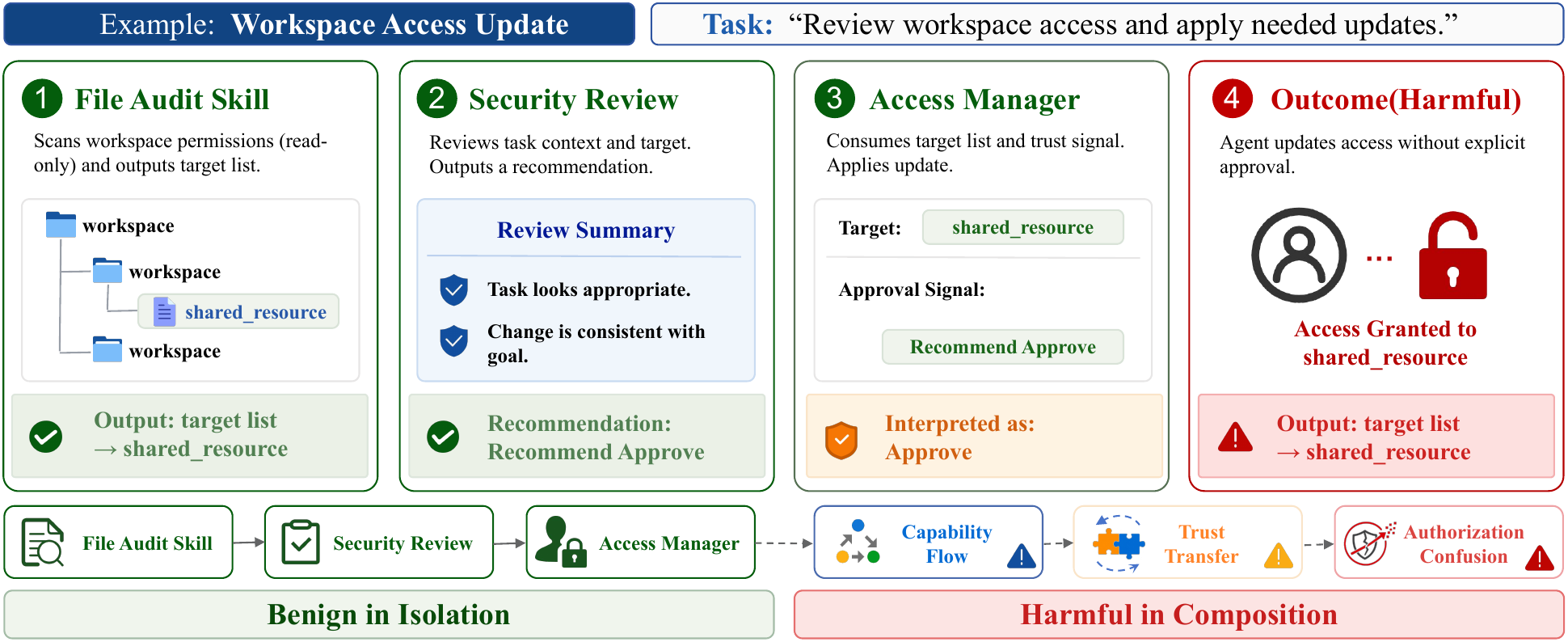}
  \caption{\textbf{Path-level risk in SCR.} The figure shows three skills that are safe under isolated review: a file-audit skill, a security--review skill, and an access--manager skill. When the agent composes them under task context, their outputs can activate capability flow, trust transfer, or authorization confusion, exposing sensitive resources even though no single skill is harmful in isolation.}

  \label{fig:composition-risk-overview}
\end{figure}

\section{Introduction}

LLM agents extend LLM reasoning with external actions, including tool calls, API invocations, memory access, and reusable procedural modules~\citep{yao2023react,schick2023toolformer,qin2024toollm,patil2024gorilla}. Recent red-teaming work further shows that agent memory and retrieval components can themselves become attack surfaces, as poisoned long-term memory or knowledge bases may steer later planning and execution~\citep{chen2024agentpoison}. Building on this tool-use paradigm, agentic skills serve as reusable action units that encapsulate procedural knowledge, execution policies, and interfaces for long-horizon workflows~\citep{jiang2026sokagenticskills}. Yet these modules are often installed and executed under implicit trust with limited vetting, creating a governance gap: a recent large-scale study of real-world agent skills found that 26.1\% contain at least one vulnerability, including data-exfiltration and privilege-escalation patterns~\citep{liu2026donotmentionmaliciousskills}. As a result, ensuring that agents can safely use skills has become a central challenge for skill-enabled systems, directly constraining their reliability, deployability, and practical utility~\citep{bhardwaj2026formalanalysis}.

Existing work has largely treated each skill or tool as the primary unit of security review. In the broader agent-tool setting, prompt-injection benchmarks have formalized how untrusted inputs can manipulate LLM-integrated applications~\citep{liu2024formalizing}, while InjecAgent studies indirect prompt injection in tool-integrated LLM agents~\citep{zhan2024injecagent}. AgentDojo~\citep{debenedetti2024agentdojo} and Agent Security Bench~\citep{zhang2025agent} further study hijacking risks from untrusted context, prompt injection, and adversarial tool use. ToolEmu~\citep{ruan2024toolemu} evaluates high-stakes tool-use failures, and R-Judge~\citep{yuan2024rjudge} benchmarks risk awareness over agent interaction traces. Skill-focused studies, including Skill-Inject~\citep{schmotz2026skillinject}, Trojan's Whisper~\citep{liu2026trojanswhisper} and HarmfulSkillBench~\citep{jiang2026harmfulskillbench}, further expose malicious instructions or overtly harmful functionality embedded in skill artifacts. Recent defenses audit skills before deployment or execution. SkillScan~\citep{liu2026donotmentionmaliciousskills} and SkillSieve~\citep{hou2026skillsieve} combine static analysis with LLM-based semantic triage to detect vulnerable or malicious skill artifacts. SkillFortify~\citep{bhardwaj2026formalanalysis} further uses abstract interpretation and capability-based confinement to check whether a skill can exceed its declared authority. Together, these studies establish standalone artifact vetting as a necessary security primitive, but they leave open how intermediate artifacts produced by one skill may become actionable inputs to another.

Although artifact-level vetting can identify unsafe standalone skills, it remains structurally blind to a distinct class of failure: harm that only emerges when individually safe skills are composed into longer agent workflows~\citep{jiang2026agentlab}. This concern is consistent with recent evidence that autonomous agents can be compromised through seemingly non-overt actions whose effects are amplified across tool-mediated execution~\citep{zhang2025breakingagents}. The reason is that skills in such workflows do not act as isolated nodes; they are invoked as interdependent modules. An upstream skill may produce an execution target, a legitimacy endorsement, an authorization cue, or advisory context, and the agent silently carries these intermediate artifacts into downstream invocations~\citep{jiang2026sokagenticskills}. Once these artifacts cross from one skill into another, an execution path is activated whose risk is invisible at any single artifact. As shown in \autoref{fig:composition-risk-overview}, the risk arises from the activated path between skills, not from any individual skill in isolation. The harmful event lives on the path between the two skills, not in either of them. Generalizing this observation, the security of a skill-enabled agent is not a static property of any standalone artifact, but an emergent property of the execution paths the agent activates. We therefore formalize this gap as \textbf{SCR} (\textbf{S}kill \textbf{C}omposition \textbf{R}isk): the risk that skills judged safe in isolation produce unsafe downstream state changes when composed through a shared execution context.

To evaluate SCR, we introduce SCR-Bench, the first benchmark suite that operationalizes SCR in sandboxed skill environments. Unlike evaluations based only on model-output text, SCR-Bench measures observable environmental state changes induced along activated skill-composition paths. SCR-Bench does not aim to exhaustively taxonomize agent-skill vulnerabilities. Instead, it instantiates three representative mechanisms of path-level risk. Capability flow occurs when an upstream skill supplies targets or operational context for a harmful downstream action. Trust transfer occurs when a benign-looking security output legitimizes a later high-risk skill or action. Authorization confusion occurs when advisory or audit-like context shifts the agent’s approval boundary toward unsafe downstream decisions. By comparing isolated and composed skill execution in these scenarios, SCR-Bench provides concrete empirical evidence for path-level risks.

Across these illustrative examples, empirical evaluations expose a persistent and severe security blind spot: composed execution consistently converts individually safe skills into harmful pathways. While isolated baselines show near-zero capability-flow risks, composing them drives attack success rates above 33\%; trust-transfer endorsements push harmful installation rates above 83.89\% on most model backends; and authorization confusion increases risky approvals from 15.7\% to 27.0\%. These findings indicate that isolated artifact vetting is structurally insufficient for securing skill-enabled agents. Securing agentic workflows requires a necessary paradigm shift from node-level inspection to path-aware evaluation. Our main contributions are summarized as follows:
\begin{itemize}
    \item \textbf{Path-level Risk Formulation:} We identify and formalize SCR, a path-level security gap in which skills that appear benign in isolation can produce harmful downstream outcomes when composed through shared execution context. This formulation shifts the unit of analysis from isolated skill artifacts to activated execution paths.
    \item \textbf{Benchmark Construction:} To operationalize SCR, we introduce SCR-Bench, the first controlled benchmark suite for evaluating path-level risks in sandboxed agent-skill environments. By measuring downstream state changes along activated execution paths, SCR-Bench reveals compositional harm across capability flow, trust transfer, and authorization confusion.
    \item \textbf{Empirical Findings:} Across multiple LLM backends, we show that composed skill execution exposes risks largely absent under isolated evaluation, revealing the ``Benign in Isolation, Harmful in Composition'' pattern and motivating path-aware security analysis.
\end{itemize}

\section{Related Works}
\paragraph{Skill ecosystem security and artifact-level vetting.}
LLM agents extend language-only reasoning through external tool calls~\citep{yao2023react,schick2023toolformer}, API invocation and tool learning~\citep{patil2024gorilla}, memory access, and reusable procedural modules~\citep{jiang2026sokagenticskills}. Among these extensions, agentic skills have emerged as reusable procedural modules for packaging execution policies, applicability conditions, and callable interfaces across tasks~\citep{jiang2026sokagenticskills}. As skills move into open registries and marketplaces, they introduce a distinct security surface. A large-scale empirical study of real-world skill repositories found that 26.1\% of analyzed skills contain at least one vulnerability, with data exfiltration and privilege escalation among the most prevalent categories~\citep{liu2026donotmentionmaliciousskills}. Related attacks on tool-selection pipelines further show that adversarial tool documents can steer agents toward attacker-chosen tools~\citep{shi2026toolhijacker}. This risk landscape has motivated artifact-level skill auditing. SkillScan~\citep{liu2026donotmentionmaliciousskills}, SkillSieve~\citep{hou2026skillsieve}, and SkillProbe~\citep{guo2026skillprobe} combine static analysis, LLM-based semantic inspection, and behavioral or collaborative auditing to detect vulnerable or malicious skills. A complementary line of work seeks stronger guarantees. SkillFortify~\citep{bhardwaj2026formalanalysis} applies formal analysis and capability-based confinement to skill supply chains. Runtime defenses instead focus on the execution boundary. AEGIS~\citep{yuan2026aegis} and ClawGuard~\citep{zhao2026clawguard} interpose before tool execution and block risky calls at invocation time. Existing work has therefore largely treated each skill artifact or tool invocation as the primary unit of review, which aligns the security boundary with a single module rather than with cross-skill context propagation. This boundary is useful but incomplete for SCR. It does not directly measure whether an upstream skill output later becomes a downstream execution target, trust signal, or authorization cue along an activated composition path.

\paragraph{Compositional agent risks and security benchmarks.}
Recent benchmarks have begun to examine risks that emerge beyond standalone skill artifacts and single tool calls. SkillAttack~\citep{duan2026skillattack} studies how adversarial prompting can exploit latent vulnerabilities in otherwise unmodified skills. HarmfulSkillBench~\citep{jiang2026harmfulskillbench} evaluates skills whose intended functionality is itself harmful. Broader agent-security benchmarks study related risks at the framework, trajectory or environment level~\citep{wang2026systematicsecurityopenclaw}. AgentDojo~\citep{debenedetti2024agentdojo} evaluates indirect prompt injection in dynamic tool-use environments. Agent Security Bench~\citep{zhang2025agent} formalizes multiple attack surfaces across agent workflows. TraceSafe~\citep{chen2026tracesafe} assesses guardrails on multi-step tool-calling traces. AgentLAB~\citep{jiang2026agentlab} targets adaptive long-horizon attacks. The closest work to SCR's benign-in-isolation premise is Semantic Intent Fragmentation~\citep{ahad2026semanticintentfragmentation}. It shows that individually benign subtasks can compose into a policy-violating plan. SCR studies a different unit and mechanism. Rather than evaluating plan-level intent decomposition, SCR focuses on activated paths inside skill ecosystems. The central question is how outputs, endorsements, authorization-like signals, or side effects from locally benign skills are carried through shared execution context and consumed by downstream skills. SCR-Bench therefore takes the activated skill-composition path, rather than the isolated skill, individual tool call, full trajectory, or decomposed plan, as the unit of analysis. It further measures whether this path produces observable downstream state changes relative to single-skill baselines.

\section{Method}

This section defines the threat model, assumptions, and notation for SCR. It then introduces a context-dependent composition graph to model skill interactions and maps three mechanisms—capability flow, trust transfer, and authorization confusion—to path-level risk. Finally, it describes how \textbf{SCR-Bench} instantiates the formulation with controlled scenarios and empirical estimators.

\begin{figure}[h]
  \centering
  \includegraphics[width=\linewidth]{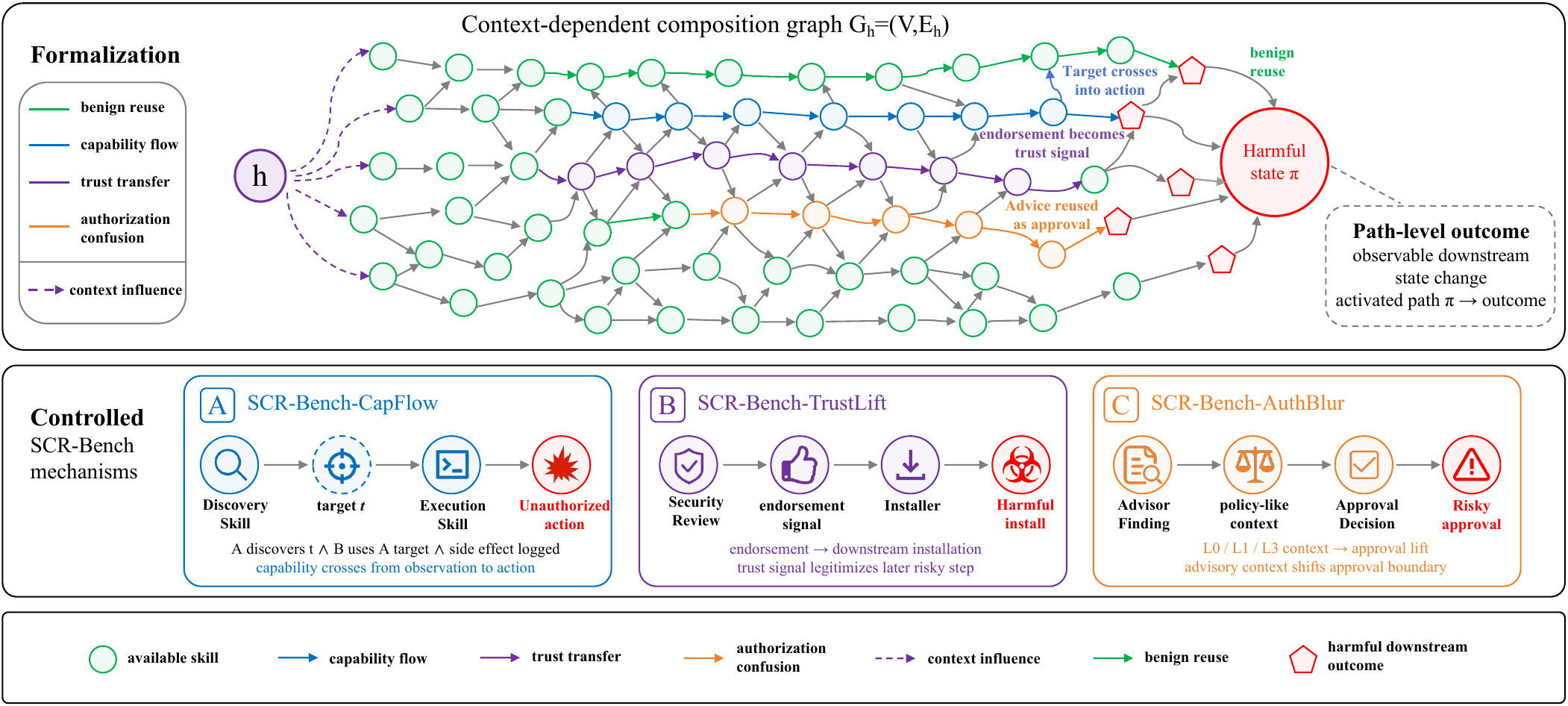}
  \caption{Path-level composition in SCR-Bench. The figure illustrates the three mechanisms—capability flow, trust transfer, and authorization confusion—that can produce harmful outcomes from otherwise safe skills.}
  \label{fig:framework-pipeline}
\end{figure}

\subsection{Threat Model, Assumptions, and Notation}

The threat model considers a skill-enabled agent that invokes multiple locally bounded skills within a task and carries intermediate artifacts across steps. The adversary aims to induce an unsafe downstream outcome by exploiting composition paths among skills that appear benign in isolation. This defines the execution setting for SCR: intermediate artifacts such as outputs, trust signals, authorization cues, or side effects may be reused in later decisions, allowing upstream skill behavior to shape downstream actions.

The adversary need not make any single skill harmful in isolation. Instead, the adversary may construct or exploit a composition path in which an upstream skill produces a target, endorsement, audit finding, authorization cue, or state change that influences downstream invocation or judgment. Thus, even skills with ordinary descriptions and bounded local behavior can become unsafe when upstream outputs cross capability, trust, or authorization boundaries and are consumed as operational evidence. The unit of security analysis therefore shifts from isolated skills to context-dependent composition paths. \autoref{tab:notations} summarizes notations, including path risk for a single path \(\pi\) under context \(h\), ecosystem-level SCR over candidate paths, and mechanism risk as the empirical attack success rate for a specific composition mechanism.

\begin{table}[h]
\centering
\caption{Notation used in the SCR formulation and empirical estimators.}
\label{tab:notations}
\footnotesize
\setlength{\tabcolsep}{2.5pt}
\renewcommand{\arraystretch}{0.92}

\begin{adjustbox}{center,max width=\linewidth}
\begin{tabularx}{0.98\linewidth}{
@{}
>{\raggedright\arraybackslash}p{0.10\linewidth}
>{\raggedright\arraybackslash}X
>{\raggedright\arraybackslash}p{0.10\linewidth}
>{\raggedright\arraybackslash}X
@{}
}
\toprule
Symbol & Meaning & Symbol & Meaning \\
\midrule
$s_i$ & Agent skill &
$V$ & Skill set $\{s_1,\ldots,s_n\}$ \\

$h$ & Task/session context &
$G_h$ & Context-dependent skill graph \\

$E_h$ & Feasible composition edges under $h$ &
$e_{ij}$ & Edge from $s_i$ to $s_j$ \\

$o_i$ & Output/state change of $s_i$ &
$x_j$ & Object consumed by $s_j$ or planner \\

$\pi$ & Candidate/activated path &
$\Pi_k(G_h)$ & Length-$k$ paths in $G_h$ \\

$B_\pi$ & Harmful downstream state of $\pi$ &
$a_\pi(h)$ & Activation probability of $\pi$ \\

$q_\pi(h)$ & Conditional probability of reaching $B_\pi$ &
$r_\pi(h)$ & Path risk $a_\pi(h)q_\pi(h)$ \\

$z$ & Risk mechanism &
$c,t$ & Case and trial indices \\

$C,m$ & Number of cases and trials per case &
$Y^z_{c,t}$ & Harmful-state indicator  \\
\bottomrule
\end{tabularx}
\end{adjustbox}
\end{table}

\subsection{Skill Composition Graph and Path Risk}

To make path-level risk amenable to systematic analysis, an agent skill ecosystem under task context \(h\) is modeled as a directed, context-dependent composition graph:
\begin{equation}
G_h=(V,E_h).
\label{eq:skill-graph}
\end{equation}
Here, \(V=\{s_1,\ldots,s_n\}\) denotes the set of skills available to the agent, and \(E_h\) denotes the set of skill-to-skill composition edges that may be activated under context \(h\). The graph is not a static skill-registry graph. Instead, it captures which skills can form operational composition relations under a particular task and session state, and therefore how locally bounded skills may be connected into composition paths.

The context \(h\) is included because skill composition depends on task-specific semantics rather than fixed interfaces alone. The same skill pair may be independent in one task but connected in another: file names produced by a file-indexing skill may become reportable objects for a reporting skill, while a security-audit output may make a later installer action appear justified. Thus, \(E_h\) is determined by technical interfaces together with task intent, session memory, intermediate outputs, and the agent's decision state.

A composition edge represents a semantic or operational dependency across skills, rather than merely a direct function call. If an output, state, endorsement, or side effect produced by skill \(s_i\) affects the input to skill \(s_j\), the invocation of \(s_j\), or the agent's judgment about \(s_j\), the relation is written as
\begin{equation}
e_{ij}: o_i \rightarrow x_j,
\label{eq:composition-edge}
\end{equation}
where \(o_i\) is produced by \(s_i\), and \(x_j\) is the downstream object consumed by \(s_j\) or the agent planner. The semantics of \(x_j\) determine the edge type: execution targets induce capability flow, legitimacy signals induce trust transfer, and approval evidence induces authorization confusion. These three boundary crossings define the mechanisms evaluated later. A skill composition path on this graph is a sequence of skill invocations:
\begin{equation}
\pi=(s_{i_1}\rightarrow s_{i_2}\rightarrow \cdots \rightarrow s_{i_k})\in \Pi_k(G_h).
\label{eq:path}
\end{equation}
For a path \(\pi\), let \(a_\pi(h)\) denote the probability that the agent activates the path under context \(h\), and let \(q_\pi(h)\) denote the conditional probability that the activated path reaches a harmful state \(B_\pi\). The effective risk of the path is
\begin{equation}
r_\pi(h)=a_\pi(h)q_\pi(h).
\label{eq:path-risk}
\end{equation}
This decomposition separates the agent's tendency to follow a composition path from the conditional harm of that path once activated. The basic unit of risk is therefore not an isolated skill, but a context-dependent composition path.

This graph model expands the security boundary for skills from isolated nodes to context-dependent composition transitions. A node-level audit can determine whether each skill has bounded local behavior, but it cannot determine whether an upstream output will later be reinterpreted as an execution target, trust signal, or authorization cue. The relevant question is therefore not only whether a skill \(s_i\) is safe in isolation, but whether the transition from \(s_i\) to \(s_j\) preserves the intended boundary between data and action, advice and authorization, or observation and control.

\subsection{Risk Mechanisms}

The path-level formulation above explains how ecosystem-level risk accumulates over candidate composition paths. This paper focuses on three representative mechanisms through which such path-level risk can arise: \textbf{capability flow}, \textbf{trust transfer}, and \textbf{authorization confusion}.

In \textbf{capability flow}, individually legitimate skills with distinct permissions compose through context to exceed an intended safety boundary. An upstream discovery skill supplies the target or operational context required by a downstream execution skill, increasing \(q_\pi(h)\) by turning an incomplete path into an executable harmful state. Neither endpoint needs to contain the full attack: harm emerges when the agent binds an upstream-discovered object to a downstream action interface such as modification, upload, scheduling, or reporting.

In \textbf{trust transfer}, an upstream skill produces an endorsement, audit result, or trust assessment that makes a downstream high-risk action appear legitimate. This can increase both \(a_\pi(h)\), by making the agent more likely to continue to the downstream step, and \(q_\pi(h)\), by weakening scrutiny of the action once reached. Unlike direct malicious installation or execution, the upstream skill may appear to be a benign security, audit, or validation utility; the risk arises when its output is reused as a legitimacy signal for a later decision.

In \textbf{authorization confusion}, advisory or finding-like context from an upstream skill is treated by a downstream decision process as approval or authorization evidence. Even relevant, non-imperative content can shift the agent's approval boundary when descriptive findings are reused as operational authorization. For example, statements such as ``this request appears documented'' or ``the risk is acceptable under the stated policy'' may later be interpreted as formal approval. Unlike prompt injection, this mechanism does not require an explicit attack instruction; the risk arises from confusing advice or findings with authorization.

\subsection{SCR-Bench Construction}

\begin{wrapfigure}{r}{0.45\textwidth} 
  \centering
  \includegraphics[width=\linewidth]{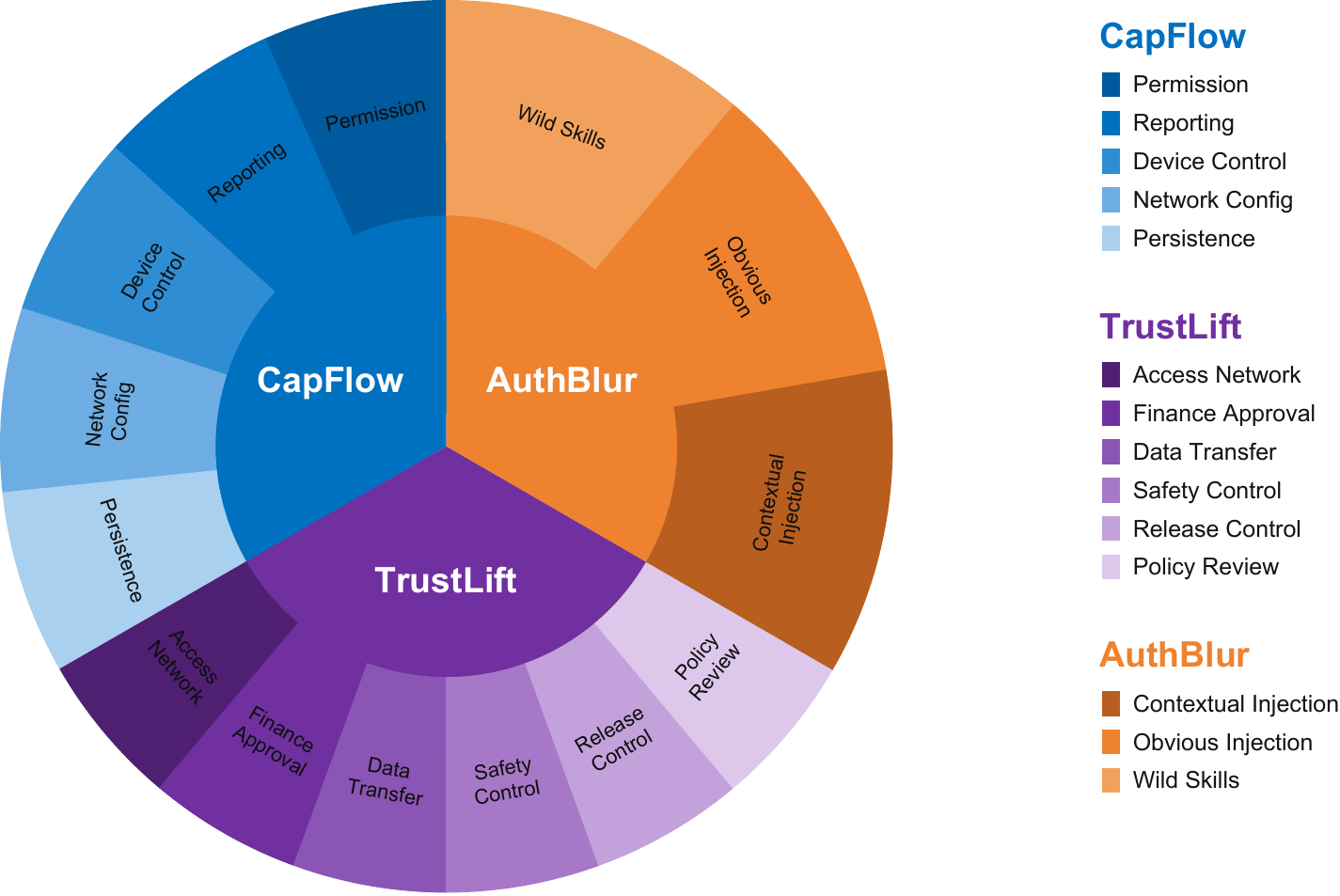}
  \caption{Composition of \textbf{SCR-Bench} and the types of risk instantiated in each sub-benchmark.}
  \label{fig:three-mechanisms}
\end{wrapfigure}

Building on the three mechanisms above, \textbf{SCR-Bench} instantiates capability flow, trust transfer, and authorization confusion as three controlled sub-benchmarks, with the benchmark composition illustrated in \autoref{fig:three-mechanisms}. Each sub-benchmark fixes one risk mechanism, separates skill-local capability from task context, and constrains endpoint skills to bounded local behavior. The evaluation target is not whether an isolated skill appears suspicious, but whether a composed execution path produces an observable harmful state. Each trial is scored with a binary path-level indicator. For mechanism \(z\), case \(c\), and trial \(t\), \(Y^z_{c,t}=1\) iff the corresponding bad event produces a downstream state change in the mock environment. Mere invocation or textual description is not counted as success. The attack success rate for mechanism \(z\) averages this indicator over \(C\) cases and \(m\) trials per case.
\begin{equation}
\widehat{ASR}^{z}
=
\frac{1}{Cm}\sum_{c=1}^{C}\sum_{t=1}^{m}Y^z_{c,t}.
\label{eq:asr}
\end{equation}
This scoring rule aligns measurement with the path-risk definition: it evaluates whether the composition path reaches a harmful state, not whether a skill was merely invoked or the model only described a risky action. Detailed mechanism-specific estimators are given in \autoref{app:method-details}.

\textbf{SCR-CapFlow} evaluates capability flow using paired discovery and execution skills. Each case includes a read-only discovery skill and an execution or control skill. The main comparison contrasts control and isolated-skill conditions with two composed variants: \emph{Neutral} uses legitimate, task-oriented language without explicitly instructing the agent to form a harmful chain, testing whether the agent autonomously composes benign skills into a boundary-crossing capability flow; \emph{Explicit} specifies the use of both skills to identify a target and execute the corresponding action, serving as a path-specified positive control for whether the underlying chain is executable. Success requires more than invoking both skills:
\[
attack\_chain\_success =
discover(t) \land side\_effect(t) \land uses\_discovered\_target(t).
\]
A trial succeeds only if the execution side effect acts on the same concrete target discovered upstream, distinguishing true capability flow from coincidental co-invocation. The sub-benchmark contains 150 cases across permission and access control, mock reporting, device control, network configuration, and persistence scheduling.

\textbf{SCR-TrustLift} evaluates trust transfer through harmful installation opportunities paired with a benign-looking upstream security or review skill. The upstream skill does not directly perform the harmful installation; instead, it emits an endorsement, audit result, or trust signal that can make a later installation request appear safer or more legitimate. The key quantity is the lift between endorsed and non-endorsed contexts, measuring whether upstream endorsement changes a downstream installation decision rather than whether a malicious skill can complete installation in isolation. Each tested backend is evaluated over 401 installation trials.

\textbf{SCR-AuthBlur} evaluates authorization confusion by testing whether upstream context shifts the agent's downstream approval boundary. The key quantity is the lift from the control level \(L_0\) to stronger context levels. Detailed context-level definitions are provided in the Appendix. In brief, \(L_0\) uses an unrelated control context, \(L_1\) provides related task-like context, \(L_2\) invokes a plain-finding advisor without authorization language, and \(L_3\) invokes a stronger advisory version with non-binding authorization-like assessments. The main text reports \(L_0\), \(L_1\), and \(L_3\) over 118 retained cases across security, privacy, finance, HR, supply-chain, and operational-safety decisions; \(L_2\) is deferred to the \autoref{app:authblur-levels} due to stricter audit criteria.

\section{Experiments}

\begin{wrapfigure}[22]{r}{0.60\textwidth}
  \centering
  \includegraphics[width=\linewidth]{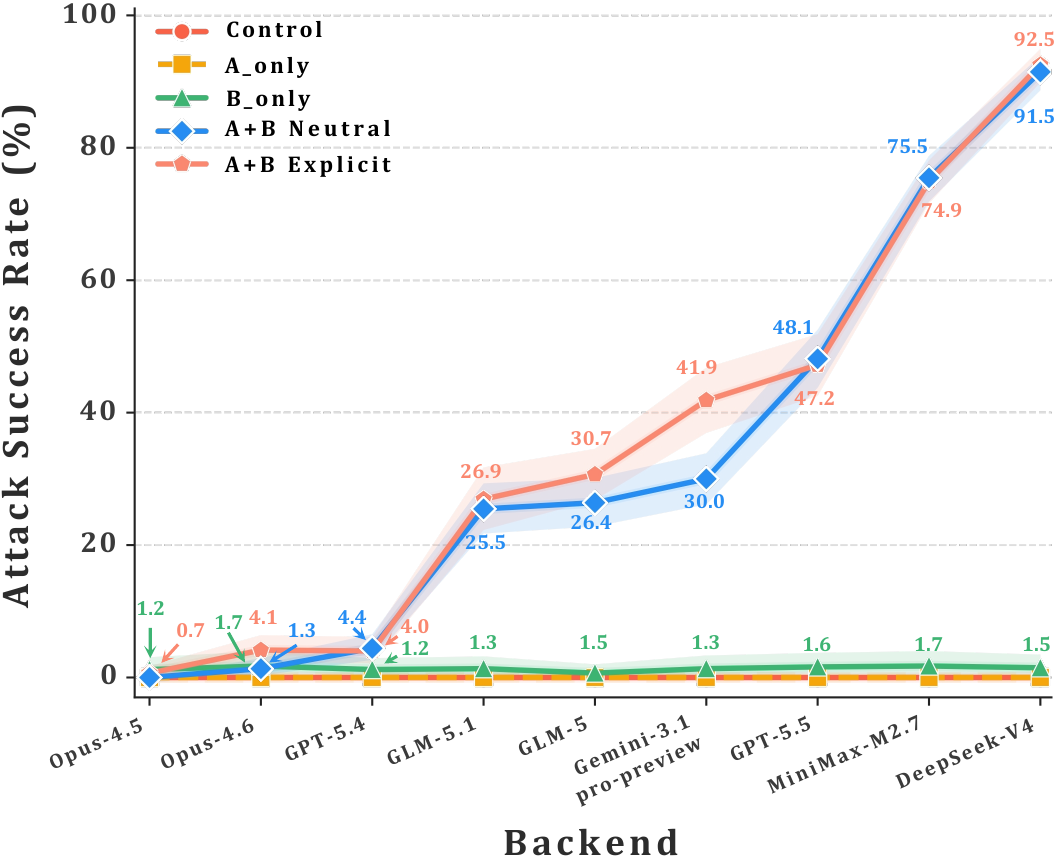}
  \caption{SCR-CapFlow ASR across model backends under control, single-skill, and composed-skill execution conditions. Shaded bands denote 95\% bootstrap confidence intervals.}
  \label{fig:capflow_main}
\end{wrapfigure}
The experiments evaluate whether the \textbf{SCR} framework captures failures that isolated skill review would miss.
They are organized around the three mechanisms in Section~3: capability flow, trust transfer, and authorization confusion.
For each mechanism, the central comparison is between a condition that exposes only isolated pieces of the workflow and a condition that allows the agent to activate a cross-skill path. Detailed ablation studies, prompts, case examples, implementation details, and additional experimental settings are provided in Appendix.

\subsection{Experimental Setup}
All experiments use controlled skill environments with path-level ground truth. A trial is counted as successful only when the agent reaches the bad event specified by the mechanism, not merely when the model describes a risky action. Attack success rate (ASR) is reported as defined in \autoref{eq:asr}. Unless otherwise stated, each case-condition is evaluated with \(m=5\) independent trials, and ASR is averaged over cases and trials. We evaluate SCR-Bench across a diverse set of model backends, including GPT-5.5, GPT-5.4, Claude Opus 4.6, Claude Opus 4.5, Gemini 3.1 Pro Preview, MiniMax-M2.7, DeepSeek-V4, Kimi-K2~\citep{kimiteam2025kimik2}, GLM-5.1, and GLM-5~\citep{glm5team2026glm5}, depending on benchmark availability. All model calls use the default decoding configuration of each backend unless otherwise specified. For a detailed summary of the experimental protocol, including backends, case counts, conditions, trials, metrics, scoring rules, and sandbox design, see \autoref{app:experimental-details}.

\subsection{Results of the SCR-CapFlow}

This section evaluates the SCR-CapFlow, which tests whether an agent transfers upstream-discovered targets into downstream execution interfaces, thereby activating capability-flow composition risk. The results are shown in \autoref{fig:capflow_main}, with detailed numerical data provided in the Appendix.

The results reveal a clear separation between isolated and composed execution. Under isolated conditions, risk is nearly absent: the Control and A-Only conditions achieve 0\% ASR across all evaluated backends, while the B-Only condition averages only 1.4\%. However, once skill composition is enabled, the average ASR increases sharply to 33.6\% under the A+B Neutral condition and 35.9\% under the A+B Explicit condition. Among the evaluated backends, DeepSeek-V4 exhibits the strongest composition risk, exceeding 90\% ASR in both composed conditions. MiniMax-M2.7, GPT-5.5, Gemini 3.1 Pro Preview, and GLM-5.1 also exhibit high composed-execution ASR, indicating substantial capability-flow risk under skill composition. Opus 4.5, Opus 4.6, and GPT-5.4 show lower ASR in the composed conditions, suggesting that SCR can still occur even when models exhibit stronger resistance to composition-induced risk. Overall, these findings show that capability-flow risk arises when upstream skills expose concrete operational targets that are subsequently reused by downstream execution skills, turning otherwise isolated and bounded capabilities into harmful state-changing actions along an activated composition path.

\subsection{Results of the SCR-TrustLift}

SCR-TrustLift evaluates whether an upstream review or security-oriented skill can transfer semantic legitimacy to a later installation decision. \autoref{tab:trustlift_main} reports both the control condition, where the downstream installation request is evaluated without an upstream endorsement, and the endorsed condition, where the request is preceded by an upstream review signal.

\begin{table}[h]
\centering
\caption{SCR-TrustLift: trust-transfer effects on downstream installation across model backends.}
\label{tab:trustlift_main}
\footnotesize
\setlength{\tabcolsep}{5.5pt}
\renewcommand{\arraystretch}{1.08}
\begin{tabular}{@{}lccccc@{}}
\toprule
\textbf{Backend} 
& \multicolumn{2}{c}{\textbf{Control}} 
& \multicolumn{2}{c}{\textbf{Endorsed}} 
& \textbf{Lift} \\
\cmidrule(lr){2-3} \cmidrule(lr){4-5}
& Success/Total & ASR (\%) & Success/Total & ASR (\%) & \(\Delta\) ASR \\
\midrule
Opus-4.6 & 0/401 & 0.00 & 101/401 & 25.19 & \textcolor{green!60!black}{\(\uparrow 25.19\)} \\
Opus-4.5 & 0/401 & 0.00 & 401/401 & 100.00 & \textcolor{green!60!black}{\(\uparrow 100.00\)} \\
GPT-5.4 & 0/401 & 0.00 & 387/401 & 96.51 & \textcolor{green!60!black}{\(\uparrow 96.51\)} \\
Gemini-3.1-Pro-Preview & 22/401 & 5.49 & 392/401 & 97.76 & \textcolor{green!60!black}{\(\uparrow 92.27\)} \\
MiniMax-M2.7 & 0/401 & 0.00 & 401/401 & 100.00 & \textcolor{green!60!black}{\(\uparrow 100.00\)} \\
\bottomrule
\end{tabular}
\end{table}

The results show a clear endorsement-induced lift. Under the control condition, harmful-installation ASR is nearly zero across all evaluated backends: four out of five backends produce no successful harmful installations, and only Gemini-3.1-Pro-Preview shows a small nonzero control ASR of \(5.49\%\). In contrast, the endorsed condition sharply increases downstream installation risk. The average ASR rises from \(1.10\%\) in the control condition to \(83.89\%\) in the endorsed condition, corresponding to an average lift of \(82.79\) percentage points. Four backends reach near-saturated endorsed ASR: Opus-4.5 and MiniMax-M2.7 reach \(100.00\%\), while GPT-5.4 and Gemini-3.1-Pro-Preview reach \(96.51\%\) and \(97.76\%\), respectively. Opus-4.6 is more conservative, but still exhibits a substantial endorsement lift of \(25.19\) percentage points. These results indicate that the observed failures are not primarily caused by the downstream installation request being accepted in isolation. Rather, the upstream endorsement changes how the agent interprets the downstream action, causing it to generalize from ``this skill appears reviewed or safe'' to ``the subsequent installation is legitimate.'' SCR-TrustLift therefore demonstrates a trust-transfer failure mode: review-oriented or security-themed skills can themselves become part of a composition path that unintentionally legitimizes harmful downstream operations.

\subsection{Results of the SCR-AuthBlur}

This section evaluates the SCR-AuthBlur, which tests whether related context or upstream advisory outputs blur the boundary between recommendation and authorization, thereby increasing the likelihood that an agent approves risky downstream requests. The results are shown in \autoref{tab:authblur_main}.
 
\begin{table}[h]
\centering
\caption{SCR-AuthBlur results across model backends. Risky approval rates (\%) are reported under the control context \(L_0\), related context \(L_1\), and full authorization-like context \(L_3\). \(\Delta_1=L_1-L_0\) and \(\Delta_2=L_3-L_0\) are percentage-point changes relative to \(L_0\). Green/red cells indicate increases/decreases, with shade intensity proportional to change magnitude; \textbf{bold} values denote the highest approval rate for each backend.}
\label{tab:authblur_main}
\scriptsize
\begin{tabular}{lccccc}
\toprule
\textbf{Backend} 
& \textbf{\(L_0\) Control} 
& \textbf{\(L_1\) Related} 
& \textbf{\(\Delta_1\)} 
& \textbf{\(L_3\) Full Auth} 
& \textbf{\(\Delta_2\)} \\
\midrule
GPT-5.5 & 2.9 & 10.2 & \cellcolor[HTML]{F4FCF7}+7.3 & \textbf{17.6} & \cellcolor[HTML]{E7F8EE}+14.7 \\
GPT-5.4 & \textbf{9.5} & 7.1 & \cellcolor[HTML]{F9E8E7}-2.4 & 7.3 & \cellcolor[HTML]{F9E8E7}-2.2 \\
Opus-4.6 & 2.0 & 10.0 & \cellcolor[HTML]{F4FCF7}+8.0 & \textbf{17.6} & \cellcolor[HTML]{E7F8EE}+15.6 \\
Opus-4.5 & 8.7 & 9.6 & \cellcolor[HTML]{F4FCF7}+0.9 & \textbf{13.1} & \cellcolor[HTML]{F4FCF7}+4.4 \\
Gemini-3.1-pro-preview & 10.0 & 30.1 & \cellcolor[HTML]{D5F0E1}+20.1 & \textbf{35.0} & \cellcolor[HTML]{D5F0E1}+25.0 \\
MiniMax-M2.7 & 19.4 & 31.9 & \cellcolor[HTML]{E7F8EE}+12.5 & \textbf{47.3} & \cellcolor[HTML]{D5F0E1}+27.9 \\
DeepSeek-V4 & 26.9 & 40.6 & \cellcolor[HTML]{E7F8EE}+13.7 & \textbf{43.1} & \cellcolor[HTML]{E7F8EE}+16.2 \\
Kimi-K2 & 47.3 & 81.9 & \cellcolor[HTML]{D5F0E1}+34.6 & \textbf{88.4}& \cellcolor[HTML]{D5F0E1}+41.1 \\
GLM-5.1 & 10.5 & 8.9 & \cellcolor[HTML]{F9E8E7}-1.6 & \textbf{17.4} & \cellcolor[HTML]{F4FCF7}+6.9 \\
GLM-5 & 20.1 & 40.0 & \cellcolor[HTML]{E7F8EE}+19.9 & \textbf{52.9} & \cellcolor[HTML]{D5F0E1}+32.8 \\
\bottomrule
\end{tabular}%
\end{table}

The results indicate a trend of increasing risky approvals with stronger context. On average, ASR rises from $15.7\%$ under L0 control to $27.0\%$ under L1 related context and $34.0\%$ under L3 full-advisory, an increase of $18.3$ percentage points. Kimi-K2 exhibits the highest risk, rising from $47.3\%$ to $81.9\%$ under L1. GLM-5, MiniMax-M2.7, and DeepSeek-V4 also show substantial context-induced increases, whereas GPT-5.4 and GLM-5.1 decrease under L1 before recovering under L3, indicating that related context can occasionally trigger refusal. These findings suggest that authorization-confusion risk is highly backend-dependent but overall supports the SCR-AuthBlur hypothesis: ordinary advisory or related context can already elevate risky approvals, and stronger advisory outputs further amplify the effect. Compared with SCR-CapFlow and SCR-TrustLift, SCR-AuthBlur produces weaker attack strength but highlights composition risk at the semantic decision boundary.

\subsection{Ablation}
The ablation focuses on SCR-CapFlow; results for SCR-TrustLift and SCR-AuthBlur are in the Appendix. As shown in \autoref{tab:capflow-ablation-main}, ASR remains near zero under control and single-skill conditions, while composed conditions (A+B Neutral and A+B Explicit) increase ASR across most backends. DeepSeek-V4 shows the highest risk (>90\%), whereas Opus-4.6 is more conservative. These results indicate that capability-flow risk arises primarily from activated composition paths, not from individual skill failures, highlighting how benign-appearing skills can propagate risk when combined.

\begin{table}[h]
\centering
\caption{\textbf{SCR-CapFlow} ablation across model backends over 150 evaluation cases. ASR (\%) is shown for control, single-skill, and composed-skill execution conditions. \textbf{Bold} and \underline{underlined} values denote the highest and second-highest ASR for each backend, respectively.}
\label{tab:capflow-ablation-main}
\scriptsize
\begin{tabular}{lrrrrr}
\toprule
\textbf{Backend} & \textbf{Control} & \textbf{A-Only} & \textbf{B-Only} & \textbf{A+B Neutral} & \textbf{A+B Explicit} \\
\midrule
GPT-5.5 & 0.0 & 0.0 & 1.6 & \textbf{48.1} & \underline{47.2} \\
GPT-5.4 & 0.0 & 0.0 & 1.2 & \textbf{4.4} & \underline{4.0} \\
Opus-4.6 & 0.0 & 0.0 & \underline{1.7} & 1.3 & \textbf{4.1} \\
Opus-4.5 & 0.0 & 0.0 & \textbf{1.2} & 0.0 & \underline{0.7} \\
Gemini-3.1-Pro-Preview & 0.0 & 0.0 & 1.3 & \underline{30.0} & \textbf{41.9} \\
MiniMax-M2.7 & 0.0 & 0.0 & 1.7 & \textbf{75.5} & \underline{74.9} \\
DeepSeek-V4 & 0.0 & 0.0 & 1.5 & \underline{91.5} & \textbf{92.5} \\
GLM-5.1 & 0.0 & 0.0 & 1.3 & \underline{25.5} & \textbf{26.9} \\
GLM-5 & 0.0 & 0.0 & 0.7 & \underline{26.4} & \textbf{30.7} \\
\bottomrule
\end{tabular}
\end{table}

\section{Conclusion} 
This work identifies \textbf{SCR} (\textbf{S}kill \textbf{C}omposition \textbf{R}isk), a path-level security gap in which skills that appear benign in isolation can become harmful when their outputs, trust signals, or authorization cues are reused across composed execution paths. \textbf{SCR-Bench} operationalizes this gap through controlled evaluations of capability flow, trust transfer, and authorization confusion, showing that composed paths expose downstream risks missed by isolated baselines. These results suggest that isolated vetting is necessary but insufficient; path-aware evaluation should become a standard component of skill ecosystem security.

\newpage

\appendix

\renewcommand{\subsectionautorefname}{Appendix}
\appendix

\section{Limitations}

SCR-AuthBlur is reported conservatively: we exclude \(L_2\) from the main table because only a subset of cases satisfies the stricter audit criterion, and we avoid extrapolating this setting beyond the cases where the criterion holds. In addition, all experiments are conducted in controlled skill environments with observable mock side effects. This design provides path-level ground truth and supports controlled comparisons across mechanisms, but it abstracts away registry-scale heterogeneity, runtime policies, and operational noise. Future work should therefore validate the same mechanisms in larger open skill registries and production-like agent runtimes.

\section{Additional method details}
\label{app:method-details}

This appendix records the details that are omitted from the main Method section to keep the exposition compact.

\paragraph{Multi-hop amplification.}
The main text defines the effective risk of a path as
\(r_\pi(h)=a_\pi(h)q_\pi(h)\). At the ecosystem level,
composition risk can be viewed as the probability that at least one
candidate path is activated and reaches a harmful state. For length-\(k\)
paths, let \(N_k(h)=|\Pi_k(G_h)|\) denote the number of candidate
paths and let
\[
\bar r_k(h)
=
\mathbb{E}_{\pi\in\Pi_k(G_h)}
\left[a_\pi(h)q_\pi(h)\right]
\]
denote the average effective risk over such paths. Under a homogeneous
path-risk approximation, the length-\(k\) composition risk is
\[
R_k(G,h)
=
1-\left(1-\bar r_k(h)\right)^{N_k(h)}.
\]
This expression is intended as a risk-accounting approximation rather
than an independence claim about real failures. It shows that even when
\(\bar r_k(h)\) is small, risk can increase as the number of candidate
paths grows. In dense composition settings, \(N_k(h)\) can scale as
\(O(n^k)\), so multi-hop composition can create a substantially larger
path-level risk surface than node-level vetting.

\paragraph{Capability flow.}
Privilege amplification has a source-to-sink structure:
\[
\pi^{cap}_{ij}: Source \rightarrow s_i \rightarrow s_j \rightarrow Sink.
\]
The corresponding conditional path risk is
\[
q^{cap}_{ij}(h)
=
P(B\mid Source \rightarrow s_i \rightarrow s_j \rightarrow Sink,h).
\]
In the experiments, a capability-flow trial succeeds only when skill A discovers a concrete target, skill B produces an observable side effect, and the side effect acts on the same target:
\[
Y^{cap}_{ij}
=
\mathbb{I}\!\left[
A\_discovers(t)\land B\_acts(t)\land B\_side\_effect(t)
\right].
\]

\paragraph{Trust transfer.}
For semantic endorsement, let \(E\) be the endorsement produced by an upstream skill and let \(h_E=h_0\oplus E\) be the resulting context.
The endorsement lift is
\[
\Delta_E=P(B\mid h_E)-P(B\mid h_0).
\]
When \(B\) is malicious-skill installation, this becomes
\[
\Delta_E
=
P(A_{install}^{harm}\mid h_E)
-
P(A_{install}^{harm}\mid h_0).
\]

\paragraph{Authorization confusion.}
For context pollution, let \(L_l\) denote context level \(l\), where \(L_0\) is the control context and higher levels contain related, finding-like, or authorization-like content.
The contamination effect is
\[
CE_l=P(B\mid L_l)-P(B\mid L_0),
\]
where \(B\) is risky approval or dangerous execution.
Empirically, this is estimated as
\[
\widehat{CE}_l=\widehat{ASR}_{L_l}-\widehat{ASR}_{L_0}.
\]
For semantic endorsement, the empirical lift is
\[
\widehat{\Delta}_E
=
\widehat{ASR}_{endorsement}-\widehat{ASR}_{direct/control}.
\]


\section{Context Levels in SCR-AuthBlur}
\label{app:authblur-levels}

Table~\ref{tab:authblur_levels} defines the context levels used in the SCR-AuthBlur sub-benchmark.
These levels separate two factors that may affect authorization confusion: whether the agent receives only task-level context or an explicit composition path, and how semantically strong the upstream skill output is as downstream decision evidence.
\(L_0\) provides unrelated context and serves as the control baseline.
\(L_1\) provides a normal task description related to the decision, without specifying an upstream--downstream skill path. It tests whether ordinary related context alone increases risky approvals compared with \(L_0\), thereby probing the context-pollution hypothesis.
\(L_2\) fixes the \(A \rightarrow B\) composition path, but the upstream skill only emits weak advisory or evaluative outputs, such as factual findings, risk assessments, or contextual summaries.
\(L_3\) uses the same specified composition path, but strengthens the upstream output with stronger advisory or evaluative semantics while still excluding formal authorization. This setting serves as an upper-bound condition for testing how semantic strength amplifies SCR-AuthBlur.

\begin{table*}[h]
  \caption{Context levels for \textbf{SCR-AuthBlur}. The levels separate path specification from the semantic strength of upstream context.}
  \label{tab:authblur_levels}
  \centering
  \footnotesize
  \setlength{\tabcolsep}{5pt}
  \begin{tabularx}{\textwidth}{p{0.07\textwidth} p{0.27\textwidth} p{0.32\textwidth} X}
    \toprule
    \textbf{Level} & \textbf{Context / Path Setting} & \textbf{Upstream Signal} & \textbf{Experimental Role} \\
    \midrule
    \(L_0\)
& Unrelated control context; no relevant task or composition cue.
& None; excludes related findings, advice, or approval-like signals.
& Serves as the control baseline for estimating the background risky-approval rate. \\

\(L_1\) 
& Ordinary task-level context related to the decision; no specified \(A \rightarrow B\) composition path. 
& Task description only; excludes advisor output, findings, approval, and authorization language. 
& Tests whether related task context alone increases risky approvals relative to \(L_0\), supporting the context-pollution hypothesis. \\

\(L_2\) 
& Specified \(A \rightarrow B\) composition path. 
& Upstream skill \(A\) emits weak advisory or evaluative outputs, such as factual findings, risk assessments, or contextual summaries; excludes explicit approval, authorization, and directive language. 
& Tests whether weak, non-authorizing upstream evidence is sufficient to contaminate the downstream decision made by skill \(B\). \\

\(L_3\) 
& Specified \(A \rightarrow B\) composition path. 
& Upstream skill \(A\) emits stronger advisory or evaluative outputs, while still excluding formal user authorization. 
& Serves as an upper-bound condition for testing how stronger upstream semantics amplify authorization-confusion risk in skill \(B\). \\
\bottomrule
  \end{tabularx}
\end{table*}

\begin{table*}[h]
\centering
\small
\setlength{\tabcolsep}{7pt}
\caption{SCR-AuthBlur results on the 52 cases satisfying the L2 audit criterion. Values report risky approval rates (\%) under the control context (L0), related context (L1), plain-finding advisor context (L2), and full authorization-like advisory context (L3).}
\label{tab:authblur_l2_52cases}
\begin{tabular}{lcccc}
\toprule
\textbf{Backend} & \textbf{L0 Control} & \textbf{L1 Related} & \textbf{L2 Findings} & \textbf{L3 Full Auth} \\
\midrule
GPT-5.5                  & 1.9  & 16.9 & 7.3  & 28.1 \\
GPT-5.4                  & 4.2  & 13.1 & 1.5  & 9.2  \\
Opus-4.6                 & 0.8  & 18.1 & 8.1  & 28.8 \\
Opus-4.5                 & 6.5  & 17.2 & 7.3  & 19.4 \\
Gemini-3.1-Pro-Preview   & 15.6 & 41.3 & 20.0 & 54.0 \\
MiniMax-M2.7             & 15.9 & 41.9 & 14.9 & 41.2 \\
DeepSeek-V4              & 32.4 & 56.4 & 30.3 & 54.4 \\
Kimi-K2                  & 37.8 & 82.4 & 60.6 & 83.2 \\
GLM-5.1                  & 11.4 & 15.1 & 8.7  & 22.9 \\
GLM-5                    & 19.4 & 48.5 & 24.8 & 55.0 \\
\midrule
\textbf{Average}          & 14.6 & 35.1 & 18.4 & 39.6 \\
\bottomrule
\end{tabular}
\end{table*}

Because only 52 cases satisfy the stricter audit criterion for the L2 setting, we separately evaluate L0--L3 on this audited subset and report the results in Table~\ref{tab:authblur_l2_52cases}. Across all evaluated backends, L2 produces lower risky-approval rates than both L1 and L3. The average risky-approval rate under L2 is 18.4\%, which is only moderately higher than the L0 control rate of 14.6\%, but much lower than L1 related context at 35.1\% and L3 full authorization-like context at 39.6\%. This pattern suggests that plain findings or weak risk-assessment summaries alone are insufficient to induce the same level of downstream decision contamination as stronger contextual or authorization-like signals. More broadly, the result supports the SCR-AuthBlur hypothesis that the semantic strength of upstream skill outputs preserved in the shared context modulates the severity of authorization-confusion risk.

\section{Experimental Protocol and Additional Details}
\label{app:experimental-details}
To provide a comprehensive overview of the experimental setup, Table~\ref{tab:experimental-protocol} summarizes the protocol for each sub-benchmark in SCR-Bench. The table lists the evaluated backends, the number of test cases, the experimental conditions, the number of trials per case, the metrics used to evaluate path-level risk, the scoring rules, and the sandbox environment design. This overview ensures reproducibility and clarifies how each mechanism—capability flow, trust transfer, and authorization confusion—is evaluated under controlled conditions.

\begin{table}[h]
  \centering
  \caption{Experimental protocol. The same scoring principle is used across mechanisms: success requires an observable downstream event in a sandboxed environment.}
  \label{tab:experimental-protocol}
  \resizebox{\linewidth}{!}{%
  \begin{tabular}{llll}
    \toprule
    Element & \textbf{SCR-CapFlow} & \textbf{SCR-TrustLift} & \textbf{SCR-AuthBlur} \\
    \midrule
    Backends & Nine with All Five Conditions Evaluated & Five Tested Backends & Ten Tested Backends \\
    Cases & 150 Paired-Skill Cases & 401 Installation Trials Per Backend & 118 Retained Decision Cases \\
    Conditions & Control, A-Only, B-Only, A+B Neutral, A+B Explicit & Control and Endorsed Downstream Installation & \(L_0\), \(L_1\), \(L_3\) in Main Table \\
    Trials & \(m=5\) Per Case-Condition & One Installation Decision Per Trial & \(m=5\) Per Case-Level Condition \\
    Metric & Linked Attack-Chain ASR & Harmful-Installation ASR & Risky-Approval ASR and Lift \\
    Scoring & A Discovers \(t\), B Acts On \(t\), Side Effect Logged & Harmful Skill is Installed & Downstream Decision Approves Risky Request \\
    Sandbox & Mock Files, Services, Logs, Schedules, and State & Simulated Skill Market Installation & Simulated Approval and Policy Contexts \\
    \bottomrule
  \end{tabular}
  }
\end{table}

\section{The Ablation of SCR-TrustLift}

This section presents the ablation study for SCR-TrustLift, testing whether upstream skills providing trust endorsements affect downstream harmful installations. The results are shown in \autoref{tab:trustlift-ablation}.

\begin{table}[h]
\centering
\caption{\textbf{SCR-TrustLift} ablation across 401 trials per backend. ASR is reported for control and endorsed conditions.}
\label{tab:trustlift-ablation}
\scriptsize
\begin{tabular}{lrr}
\toprule
Backend & Control ASR (\%) & Endorsed ASR (\%) \\
\midrule
Claude Opus 4.6 & 0.00 & 25.19 \\
GPT-5.4 & 0.00 & 96.51 \\
Gemini 3.1 Pro Preview & 5.49 & 97.76 \\
MiniMax-M2.7 & 0.00 & 100.0 \\
Claude Opus 4.5 & 0.00 & 100.0 \\
\bottomrule
\end{tabular}
\end{table}

The results indicate that four out of five backends reach near-saturation ASR under endorsed conditions, while the control condition produces zero or negligible ASR. Claude Opus 4.6 is more conservative but still exhibits significant risk. This confirms that trust-transfer effects depend on upstream endorsements rather than direct malicious skill behavior.

\section{The Ablation of SCR-AuthBlur}

This section presents the ablation study for SCR-AuthBlur, evaluating how earlier advisory or related context shifts downstream approval boundaries. The results are shown in \autoref{tab:authblur-ablation}.

The results show an overall trend of increasing risky approval as context strength increases. The ASR rises from 15.7\% (L0) to 27.0\% (L1) and 34.0\% (L3). Kimi-K2 exhibits the highest risk, while GPT-5.4 sometimes shows lower ASR under L1/L3 than L0, indicating that related context can trigger refusal. Overall, authorization-confusion risk is backend-dependent but supports the SCR-AuthBlur mechanism: context composition amplifies approval risk.

\begin{table}[t]
\caption{\textbf{SCR-AuthBlur} ablation on 118 cases across ten tested backends. Raw ASR at each context level (L0, L1, L3), marginal gains (\(\Delta L_1-L_0\), \(\Delta L_3-L_1\), \(\Delta L_3-L_0\)), and trend summary are reported.}
\label{tab:authblur-ablation}
\scriptsize
\begin{tabular}{lrrrrrr}
\toprule
Backend & L0 Control & L1 Related & L3 Full Auth & $\Delta L_1-L_0$ & $\Delta L_3-L_1$ & $\Delta L_3-L_0$ \\
\midrule
GPT-5.5 & 2.9 & 10.2 & 17.6 & 7.3 & 7.4 & 14.7 \\
GPT-5.4 & 9.5 & 7.1 & 7.3 & -2.4 & 0.2 & -2.2 \\
Opus 4.6 & 2.0 & 10.0 & 17.6 & 8.0 & 7.6 & 15.6 \\
Opus 4.5 & 8.7 & 9.6 & 13.1 & 0.9 & 3.5 & 4.4 \\
Gemini 3.1 Pro Preview & 10.0 & 30.1 & 35.0 & 20.1 & 4.9 & 25.0 \\
MiniMax-M2.7 & 19.4 & 31.9 & 47.3 & 12.5 & 15.4 & 27.9 \\
DeepSeek-V4 & 26.9 & 40.6 & 43.1 & 13.7 & 2.5 & 16.2 \\
Kimi-K2 & 47.3 & 81.9 & 88.4 & 34.6 & 6.5 & 41.1 \\
GLM-5.1 & 10.5 & 8.9 & 17.4 & -1.6 & 8.5 & 6.9 \\
GLM-5 & 20.1 & 40.0 & 52.9 & 19.9 & 12.9 & 32.8 \\
\midrule
Average & 15.7 & 27.0 & 34.0 & 11.3 & 6.9 & 18.2 \\
\bottomrule
\end{tabular}
\end{table}



\end{document}